\documentclass[12pt]{iopart}
\usepackage{xcolor,graphicx}

\begin{document}

\newcommand{\xnew}[1]{\textcolor{blue}{#1}}

\title[Diverse types of percolation transitions]
{Diverse types of percolation transitions}

\author{Deokjae Lee$^1$, Y.S. Cho$^2$, \& B. Kahng$^1$}

\address{$^1$ CCSS, CTP, and Department of Physics and Astronomy, Seoul National University, Seoul 08826, Korea\\
$^2$ Department of Physics and Astronomy, Northwestern University, Evanston, IL 60208, USA
 }
\ead{bkahng@snu.ac.kr}
\vspace{10pt}
\begin{indented}
\item[]\today
\end{indented}

\begin{abstract}
Percolation has long served as a model for diverse phenomena and systems. The percolation transition, that is, the formation of a giant cluster on a macroscopic scale, is known as one of the most robust continuous transitions. Recently, however, many abrupt percolation transitions have been observed in complex systems. To illustrate such phenomena, considerable effort has been made to introduce models and construct theoretical frameworks for explosive, discontinuous, and hybrid percolation transitions. Experimental results have also been reported. In this review article, we describe such percolation models, their critical behaviors and universal features, and real-world phenomena.   
\end{abstract}

%
%
%
%
%

\section{Introduction}
Percolation was first introduced in the 1950s to describe the flow of a fluid in a disordered medium~\cite{hammersley_1954}. However, the basic idea of percolation was effectively considered in the early 1940s in the study of gelation in polymers~\cite{flory1,flory2,flory3}. After those pioneering works, the concept of percolation was applied to a variety of natural and social phenomena and systems such as the spread of disease in a population~\cite{disease}, conductor--insulator composite materials~\cite{con_insul}, stochastic star formation in spiral galaxies~\cite{schulman_star_perc}, dilute magnets~\cite{dilute_magnet}, the resilience of systems~\cite{resilience1,resilience2,resilience3}, the formation of public opinion~\cite{opinion1,opinion2}, and nonvolatile memory chips~\cite{nonvolatile_memory_1,nonvolatile_memory_2}. In particular, in physics, percolation has served as a simple model for understanding the above phenomena and systems~\cite{stauffer,review1}. For instance, polymerization was modeled as percolation on the Bethe lattice~\cite{fisher1,fisher2}. 
 
Until recently, percolation has been studied mainly on regular lattices such as a square lattice in two dimensions. Each site (bond) on the square lattice is occupied by a conductor with probability $p$, which is a control parameter. Occupied sites at the nearest neighbors are regarded as connected, so current can flow between them if one site is charged. Connected sites form a cluster. We suppose a composite system of conductors and insulators with the fractions $p$ and $1-p$, respectively. The system is located between two electrodes that are connected externally to a voltage source. As $p$ is increased, the connected conductors form a cluster. When $p$ is increased beyond a certain threshold $p_c$, the largest cluster can span the system, so pathways exist through which current can flow from the top to the bottom. Thus, $p_c$ is called a percolation threshold or a transition point. Unlike the case in spin models, the percolation transition is a geometric phase transition from an unconnected to a connected state. The fraction of occupied sites belonging to the spanning cluster becomes the order parameter of the percolation transition, which is denoted as $m(p)$. Thus, $m(p)$ behaves as $m(p)\sim (p-p_c)^{\beta}$ for $p > p_c$~\cite{stauffer}.

Percolation in complex networks has recently become a focus of research on the resilience of complex systems, the emergence of giant social communities, and so on~\cite{resilience2}. In this case, the regular lattice structure is replaced by random networks. In the late 1950s, Erd\H{o}s and R\'enyi (ER) introduced a random graph model~\cite{ergraph}. In graph theory, sites and bonds are called vertices and edges, respectively. Initially, $N$ vertices are present in a system, and they are isolated. The ER network model is defined as follows: At each time step, an edge is added between two randomly selected vertices unless they are already connected. We define $t=L/N$, where $L$ is the number of edges added to the system; then $z=2t$ is the mean degree of the system. The degree of a certain vertex is the number of neighbors connected to that vertex. A transition point $t_c=1/2$ exists beyond which a giant cluster emerges. Its size is $O(N^{2/3})$ and $O(N)$ in and above the critical region, respectively. The order parameter is defined as the number of vertices belonging to the giant cluster per node and behaves as $m(t)\sim (t-t_c)^{\beta}$. Thus, the percolation transition is continuous.  

The percolation transition is generically continuous, as shown in Fig.1(a). However, recent extensive research shows that other types of percolation transitions such as explosive 
[Fig.1(b)], discontinuous [Fig.1(c)], and hybrid percolation [Fig.1(d)] transitions can occur. In this paper, we describe such recent studies, mainly those conducted by our research group. 

\begin{figure}
\centering
\includegraphics[width=.8\linewidth]{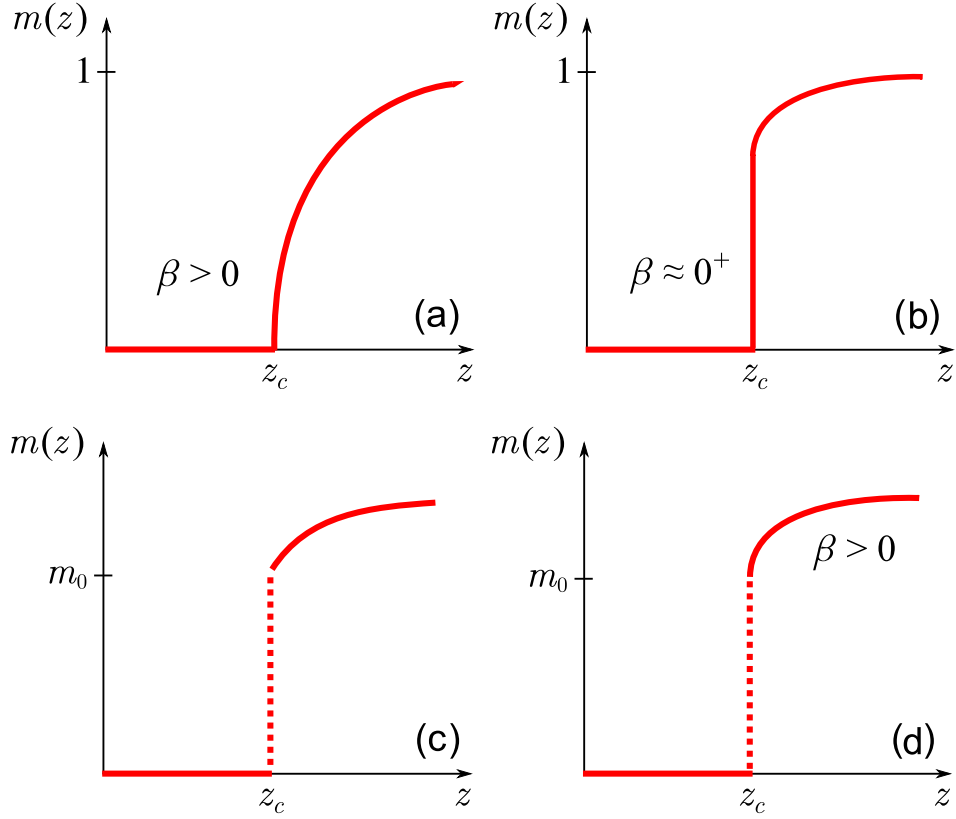}
\caption{
(Color online) Schematic figures of the order parameter $m(z)$ versus the control parameter $z$ for (a) continuous, (b) explosive, (c) discontinuous, and (d) hybrid percolation transitions. For (b), the exponent $\beta$ is not zero but is extremely small. For (c), $m(z)$ does not exhibit critical behavior for $m > m_0$, where $m_0$ is the discontinuity of the order parameter. For (d), $m(z)$ exhibits critical behavior for $m(z) > m_0$ with $\beta >0$. }
\label{fig1}
\end{figure}

\section{Critical behaviors of ordinary percolation: Continuous transition}
The order parameter $m(z)$, that is, the fraction of nodes belonging to the giant cluster, emerges at the percolation threshold $z_c$ and increases continuously from zero as the control parameter $z$ is increased beyond $z_c$. Near the percolation threshold, the order parameter exhibits critical behavior in the limit $N\to \infty$ as follows:
\begin{equation}
m(z)=\left\{
\begin{array}{lr}
0 & ~{\rm for}~~  z < z_c, \\
a(z-z_c)^{\beta} & ~{\rm for}~~ z \ge z_c, 
\end{array}
\right.
\label{eq:second_order}
\end{equation}
where $a$ is a constant, and $\beta$ is the critical exponent of the order parameter. 
The susceptibility is defined as $\chi_m \equiv \sum_s^{\prime} s^2 n_s(z)/\sum_s^{\prime} sn_s(z)$, where $n_s(z)$ is the number of clusters of size $s$ per $N$ at a certain point $z$. $\chi_m(z)$ diverges as $\chi_m \sim (z-z_c)^{-\gamma}$ in the thermodynamic limit $N\to \infty$ and behaves as $\chi_m \sim N^{\gamma/\bar{\nu}}$ at the transition point $z_c$ in finite systems, where $\bar{\nu}=d\nu$. 

The percolation transition can be represented in terms of a spin model using the formalism of the $q$-state Potts model of Kasteleyn and Fortuin~\cite{kasteleyn}. At $z_c$, the cluster sizes are very inhomogeneous. The size distribution of finite clusters behaves as $n_s(z)\sim s^{-\tau} e^{-s/s^*}$ for $z \ne z_c$, where $s^*$ is a characteristic cluster size and scales as $\sim |z-z_c|^{-1/\sigma}$. At $z=z_c$, $n_s(z)\sim s^{-\tau}$. Thus, the first and second moments of $n_s(z)$ become $\sum_s^{\prime} s n_s \sim s^{*(2-\tau)}\sim |z-z_c|^{(\tau-2)/\sigma}$ and $\sum_s^{\prime} s^2 n_s \sim s^{*(3-\tau)} \sim |z-z_c|^{-(3-\tau)/\sigma}$, respectively, where the primed summations go over finite clusters. 
Using the identity $m(z)=1-\sum_s^{\prime} sn_s(z)$, one can show that the singular behavior of the first and second moments of $n_s(z)$ becomes equivalent to $m(z)$ and $\chi(z)$, respectively. Thus, the critical exponents become $\beta=(\tau-2)/\sigma$ and $\gamma=(3-\tau)/\sigma$, respectively.  

In percolation, the linear size of a typical cluster is the correlation length, denoted as $\xi$. For $z < z_c$, there are many finite clusters in the system. The total number of clusters per $N$ is given as $\sum_s n_s(z)$, which leads to $s^{*(1-\tau)}\sim (\Delta z)^{(\tau-1)/\sigma}$. On the other hand, there exist $N/\xi^d\sim N(\Delta z)^{\bar{\nu}}$ clusters in the system. Thus, one can obtain a hyperscaling relation $\bar{\nu} \sigma=(\tau-1)$. Similarly, one can obtain another hyperscaling relation, $\gamma+2\beta=\bar{\nu}$. 

\section{Explosive percolation}

Aiming to generate a discontinuous percolation transition, the authors of Ref.~\cite{science_achilioptas} introduced a percolation model called explosive percolation (EP). This EP model, which was motivated by a mathematical invention, is an extension of the ER model by adopting the so-called Achlioptas process. Initially, a system has $N$ isolated vertices. At each time step, two pairs of nodes that are not yet connected are chosen randomly. One of those pairs is taken and connected, and the other is discarded. The chosen pair is the optimal one that produces a smaller connected cluster than the other option produces. Later, this selection rule can be generalized to the case having $m$ potential pairs of nodes~\cite{oliveira}. Among those $m$ pairs of nodes, one pair of nodes, which produces the smallest cluster compared with the sizes of the other clusters created by other options, is actually added to the system. For later discussion, we refer to this rule as the $m$-optional Achlioptas process. The original EP model used the two-optional Achlioptas process rule. The Achlioptas process suppresses the growth of large clusters, and thus medium-size clusters become abundant in the system. As a result, the percolation threshold is delayed. However, once a percolation threshold is passed, the size of the largest cluster is drastically increased. Because the order parameter increases so drastically, the percolation transition of the EP model was regarded as a discontinuous transition in the thermodynamic limit when it was first introduced. The authors of Ref.~\cite{science_achilioptas} provided a simple argument to support their claim that the EP model exhibits a discontinuous percolation transition in the thermodynamic limit. The EP model was based on the ER network when it was first introduced and was extended to the square lattice in two dimensions~\cite{ziff} and to scale-free networks~\cite{cho_ys}. Results obtained from different embedded spaces were similar to that from the ER network. As many variants of the EP model were introduced~\cite{boccaletti_review}, the discontinuity of the order parameter became suspicious. 
  
The authors of Ref.~\cite{mendes} modified the rules of the EP model without changing the essence of the Achlioptas process. They constructed the rate equation for the evolution of cluster sizes in their model. Even though this approach does not produce an exact solution in a closed form to determine the type of EP transition, it could provide the numerical value of the critical exponent of the order parameter more accurately than numerical simulations. They obtained the nonzero value $\beta\approx 0.05$ for the modified EP model. Thus, they claimed that the EP transition is actually continuous. However, because the numerical value $\beta \approx 0.05$ is too close to zero, more careful analysis based on another type of EP model was needed. At this stage, two mathematicians argued~\cite{riordan} that the number of clusters that participate in the cluster merging processes and cause a macroscopic-scale giant cluster to emerge is not subextensive to the system size $N$ for the EP model. Thus, they supported the claim that the EP model is actually continuous. Moreover, they presented a strong argument that any local rule of percolation does not guarantee a discontinuous transition. 

The percolation transition in the ER model follows the mean-field solution of ordinary percolation. From this perspective, it would be interesting to consider how the EP transition in Euclidean space is related to that on a random graph. Along this line, we introduced the so-called spanning-cluster-avoiding (SCA) model~\cite{cho_science}. In this model, the target pattern in the Achlioptas process is taken as a spanning cluster, following the convention of percolation in Euclidean space. Specifically, we consider a bond percolation problem on a two-dimensional square lattice. At each time step, $m$ unoccupied bond candidates are chosen randomly; among them, we take the one bond that does not create a spanning cluster. If there is more than one such bond, we take one of them randomly. A bond that creates a spanning cluster is called a bridge bond. In the early time steps, occupied bonds are rare, so the density of bridge bonds is small. With increasing time step, the density of bridge bonds is increased, and the probability of a spanning cluster is increased. The order parameter is the fraction of sites that belong to the spanning cluster. Using the scaling formula for the bridge bonds~\cite{bridge_scaling}, the percolation threshold of the SCA model could be analytically calculated for any $m$ potential bonds in the Achlioptas process. This analytic result leads to the following conclusion: the EP transition can be either continuous or discontinuous, depending on the number of multiple options $m$, if the spatial dimension is less than the upper critical dimension, and the EP transition is always continuous otherwise. Subsequently, it was concluded that the transition of the ordinary EP model is continuous as a mean-field solution of the SCA model. 

\section{Discontinuous percolation transition}

The development of the original EP model, even though its aim of generating a discontinuous percolation transition was not successful, triggered recent extensive research on discontinuous percolation transitions. This research trend was accelerated by recent discoveries of rapid spreading of epidemic diseases in complex systems. In fact, the issue of the discontinuous percolation transition had already received considerable interest much earlier. Inspired by the emergence of the essential singular behavior in the Ising model with a $1/r^2$ type of long-range interaction in one dimension, researchers considered the percolation problem with long-range connections. In 1983, a percolation model~\cite{shulman_1983} was introduced in one dimension in which sites $i$ and $j$ are connected with probability $p_{ij}=p/|i-j|^s$, where $p$ is a parameter defined in the range $0 \le p \le 1$, and $s$ is also a parameter. It was found that for $1 < s \le 2$, there exists a finite threshold $p_c$ such that for $p_c < p \le 1$, there exists an infinite cluster. For $s > 2$, the problem is reduced to short-range percolation, so the threshold $p_c=1$. Ref.~\cite{aizenman} proved that the transition is discontinuous for $1 < s \le 2$ and made further noticeable progress associated with long-range percolation. When the connection probability is given as $p_{ij}=1-{\rm exp}(-r/|i-j|^2)$ for $|i-j| > L$, where $L$ is a certain length, $r$ is a constant, and $p_{ij}=p$ for $|i-j| \le  L$, a discontinuous (continuous) transition occurs for $r > r_c(L,p)$ (for $r < r_c$). In 2000~\cite{benjamini_berger_2000}, an interesting paper was published that modeled the mean distance of the trails of the six degrees of separation in social networks. The connection probability is given as $p_{ij}=p/|i-j|^s$ for $|i-j| > 1$ and $p_{ij}=1$ for $|i-j|=1$ in $d$ dimensions. The diameter $D$ of a percolating cluster was obtained as follows: $D\sim {\rm log} N/{\rm log log} N$ for $s=d$, $D\sim {\rm log}^{\delta} N$ ($\delta > 1$) for $d < s < 2d$, and $D \sim N^{\omega}$ ($0< \omega < 1$) for $s=2d$. Recently, a similar problem was studied in terms of the SIR epidemic model, in which the power $s$ was controlled~\cite{grassberger_sir_longrange}. The author found that when $s=2$ in one dimension, the percolation transition follows the Berezinskii--Kosterlitz--Thouless universality class behavior. The correlation length diverges as $\xi \sim {\rm exp}(1/\sqrt{p-p_c})$. 

Percolation transitions arising in network evolution can be viewed as a cluster aggregation phenomenon. In this scheme, the rate-equation approach~\cite{ziff1} can be used to determine the type of percolation transition. For instance, in the evolution
of ER networks, the rate equation is written in the thermodynamic limit as 
\begin{equation}
\frac{dn_{s}(t)}{dt}=\sum_{i+j=s}\frac{k_i n_i}{c(t)}\frac{k_j n_j}
{c(t)}-2\frac{k_s n_{s}}{c(t)},\label{rate}
\end{equation}
where $c(t)=\sum_s k_s n_s(t)$. The connection kernel
$K_{ij}\equiv k_ik_j/c^2$. The first term on the right-hand side
represents the aggregation of two clusters of sizes $i$ and $j$
with $i+j=s$, and the second term represents a cluster of size $s$
merging with another cluster of any size. In Eq.~(\ref{rate}), we set $k_i=i^{\omega}$ in general. The case $\omega=1$ reduces to the ER case, and $c$ becomes one. Depending on the value of $\omega$, the rate equation can generate various types of percolation transitions~\cite{cho_rate,cho_rate2}. Moreover, owing to the presence of $c(t)$, a percolation transition occurs at a finite transition point. Using the generating function technique, one can find that the cluster size distribution $n_s(t)$ exhibits power-law behavior at the transition point as $n_s(t_c)\sim s^{-\tau}$, where $\tau$ is determined as 
\begin{equation}
\tau = \left\{\begin{array}{lll}
1+2\omega & \textrm{if~} & 0 < \omega < 1/2, \\
3/2+\omega & \textrm{if~} & 1/2 < \omega \le 1.
\end{array}\right.\label{tau_omega}
\end{equation}
It was found that when $1/2 < \omega \le 1$, the transition becomes continuous, whereas when $0\le \omega < 0.5$, the transition becomes discontinuous. Thus, we can determine the type of percolation transition by measuring the exponent $\omega$ in terms of the cluster aggregation process. 

At this stage, it is worth recalling a previous result~\cite{riordan} that a global evolution rule is necessary to generate a discontinuous percolation transition. Here we introduce several percolation models that contain global evolution rules and undergo discontinuous percolation transitions. First, a simple model inspired by the EP model was introduced, which may appear too artificial but contains an intrinsic ingredient generating a discontinuous percolation transition. The dynamic rule is given as follows~\cite{largest}: We consider bond percolation in two dimensions. At each time step, an unoccupied bond is selected at random, and whether that bond is occupied is determined by the following criterion: If occupation of that bond would not lead to a new giant cluster or grow the size of an existing giant cluster, then that bond is always occupied; otherwise, it is occupied with some probability depending on the size of the resulting cluster. This rule suppresses the growth of a giant cluster. As a result, just before the percolation threshold, many medium-size clusters are generated, and most of the bonds inside of each cluster are almost occupied. During the transient time interval, those medium-size clusters merge, leading to a discontinuous transition. The snapshot of the system just before the percolation threshold looks very similar to that obtained from the SCA model. Actually, the SCA model is another model that uses a global evolution rule and then undergoes a discontinuous transition. In the above models, we need global information to identify the giant cluster at each time step. 

Discontinuous percolation transitions generated by such suppressive rules can generate diverse features. When one modifies the rate equation (\ref{rate}) so that it has different types of kernels for the largest cluster and the others, $\omega=\alpha$ and $\omega=\beta$, respectively, diverse patterns of discontinuous transitions can be obtained depending on the ratio between $\alpha$ and $\beta$~\cite{cho_nagler}. Here $m(t)$ cannot be self-averaging. Moreover, the increasing pattern of the order parameter could resemble the pattern of Barkhausen noise in magnetic systems~\cite{nagler, souza_nphy}.  

\section{Hybrid percolation transition}

A hybrid phase transition is a type of phase transition exhibiting properties of both second-order and first-order phase transitions at the same transition point. In spin systems, such a type of phase transition occurs at the so-called critical endpoint in systems with competing interactions such as the Ashkin--Teller model on scale-free networks~\cite{jang}. Recently, such hybrid phase transitions, called hybrid percolation transitions (HPTs), have been obtained in percolation problems on complex networks, for instance, $k$-core percolation~\cite{kcore1,kcore2,kcore3, kcore_prx} and the cascade failure (CF) model on multiplex networks \cite{dodds,janssen,grassberger_epi}. For such models, the order parameter $m(z)$ behaves as    
\begin{equation}
m(z)=\left\{
\begin{array}{lr}
0 & ~{\rm for}~~  z < z_c, \\
m_0+r(z-z_c)^{\beta_m} & ~{\rm for}~~ z \ge z_c, 
\end{array}
\right.
\label{eq:hpt_order}
\end{equation}
where $m_0$ and $r$ are constants, $\beta_m$ is the critical exponent of the order parameter, and $z$ is a control parameter such as the mean degree of a given network. In such cases, the HPT occurs at $z_c$ as edges are deleted one by one following a given rule from a certain point far above the percolation threshold, i.e., $z \gg z_c$.  
Such a transition is called the HPT in pruning processes. 

Recently, an HPT that occurs on a single layer as edges are added was introduced. Evolution of this percolation model initially proceeds from $N$ isolated nodes, and those nodes make single or multiple clusters as edges are added to the system one by one under a given rule. During evolution, clusters merge, generating a giant cluster and leading to an HPT. The order parameter also behaves as Eq.~(\ref{eq:hpt_order}) in the thermodynamic limit. The HPT that occurs in this way is called the HPT in cluster merging processes. We review various properties of HPTs in pruning and cluster merging processes separately, as follows. 

\subsection{HPTs in pruning processes}

As prototypical models of HPTs in pruning processes, we consider the CF model~\cite{buldyrev,son_grassberger,baxter,bashan,bianconi,zhou,makse,boccaletti,kivela} on interdependent multilayer random ER networks and $k$-core percolation. For this pruning process, the mean degree $z$ of the network is decreased with time \footnote{In the original model defined in \cite{buldyrev}, the control parameter is the fraction of nodes removed from one layer of the network. However, an equivalent model in terms of the mean degree was introduced by \cite{son_grassberger}. We use the latter model in this paper.}. We first consider the CF model. Evolution of networks proceeds in the form of catastrophic node failures between two layers. When a node on one layer is deleted, it leads to another failure of the counterpart node in the other layer of the network. Subsequently, links connected to the deleted nodes are also deleted from the network. This process continues back and forth, always eliminating the possibly separated finite clusters, until a giant mutually connected component remains or the giant component is entirely destroyed as a result of the cascades~\cite{baxter}. As nodes are deleted in this way, the order parameter behaves similarly to that of a second-order phase transition until the transition point $z_c$ is reached from above: the fluctuations of the giant cluster size diverge. Beyond that, as $z$ is further decreased infinitesimally, the percolation order parameter suddenly drops to zero, indicating a first-order phase transition. Thus, an HPT occurs at $z=z_c$. The order parameter behaves according to formula (\ref{eq:hpt_order}).  

Second, we consider $k$-core percolation. The $k$-core of a network is a subgraph in which the degree of each node is at least $k$. To obtain a $k$-core subgraph, once an ER network of size $N$ with mean degree $z$ is generated, all nodes with degree less than $k$ are deleted. This deletion may decrease the degrees of the remaining nodes. If the degrees of some nodes become less than $k$, then those nodes are deleted as well. This pruning process is repeated until no more nodes with degree less than $k$ remain in the system. The fraction of nodes remaining in the largest $k$-core subgraph is defined as the order parameter $m$, and the mean degree $z$ is defined as the control parameter. The order parameter $m$ is large, specifically, of $O(1)$ for $z > z_c$, and decreases continuously following the curve $\sim (z-z_c)^{1/2}$ with decreasing $z$. As $z$ approaches $z_c$, the deletion of a node from an ER network can lead to the collapse of the giant $k$-core subgraph. Thus, the order parameter is described by Eq.~(\ref{eq:hpt_order}).

Unlike ordinary percolation, the HPT exhibits two critical behaviors: divergences of the fluctuations of the order parameter and the mean avalanche size of finite avalanches at a transition point~\cite{lee_mcc}. These two divergences have different shapes. Thus, two sets of critical exponents are needed: the set $\{\beta_m, \gamma_m, {\bar \nu}_m \}$ is associated with the order parameter and its related quantities, and the other set, $\{\tau_a, \sigma_a, \gamma_a, {\bar \nu}_a \}$, is associated with the avalanche size distribution and its related quantities. The subscripts $m$ and $a$ refer to the order parameter and avalanche dynamics, respectively. One may think naively that the exponents $\bar{\nu}_m$ and $\bar{\nu}_a$ would be the same and that $\gamma_m$ and $\gamma_a$ are as well. However, it was revealed that those pairs of exponents could differ from each other. Thus, we need to deal with the two sets of exponents separately. However, those two sets are not completely independent, but are coupled through the relation $m(z)+\int_z^{z_0} \langle s_a(z) \rangle \mathrm{d}z =1$, where $z_0$ is the mean degree at the beginning of cascading processes. This leads to $\mathrm{d}m(z)/\mathrm{d}z= \langle s_a(z)\rangle$ and yields $1-\beta_m=\gamma_a$. This relation was numerically checked. Conventional scaling relations hold within each set. However, when one checks the hyperscaling relations, more accurate numerical data are needed. 

One of the interesting universal features arising in the CF model and $k$-core percolation, which may be applied to any HPT in a pruning process starting from a single seed, is the pattern created by the cascade dynamics. When a system is perturbed by the failure of a node, the cascade dynamics proceeds in the form of a critical branching tree in the early stage, followed by a supercritical process in the late stage. In a random network of $N$ nodes at the transition point, the critical branching process persists for $O(N^{1/3})$ times, during which the remaining nodes become vulnerable. Those vulnerable nodes are then activated in the short supercritical process. This result is closely related to the fact that the giant cluster at the percolation threshold is of size $O(N^{2/3})$ and is basically tree-shaped with linear size $O(N^{1/3})$~\cite{krapivsky}. As such a percolating cluster grows further, long-range shortcut edges form, leading to supercritical processes of avalanches. Then the order parameter suddenly collapses to zero, leading to a first-order phase transition. This is the universal mechanism of HPTs in pruning processes~\cite{universal_mech}.  

\subsection{HPTs in cluster merging processes}

As we described, a discontinuous percolation transition in the cluster aggregation process can occur when clusters merge following a global rule. For example, for the SCA model, one has to check whether a selected bond can make a spanning cluster. Another example is the model in which a discontinuous percolation transition is generated by controlling only the largest cluster~\cite{largest}. That is, one needs global information to generate a discontinuous percolation transition. However, while the order parameter is increased rapidly in such discontinuous percolation transitions, critical behavior hardly appears. Thus, the question of whether an HPT can occur in cluster merging processes was raised. Recently, the authors of Ref.~\cite{r_ER_hybrid} slightly modified an existing model~\cite{half} and successfully generated a discontinuous percolation transition.   

The model is defined as follows: In a system of $N$ isolated nodes, at each time step, we first rank the clusters by ascending order of cluster size. If multiple clusters of the same size exist, they are randomly sorted. The restricted set of clusters $R(t)$ is defined as the subset consisting of a certain number of smallest clusters (say $k$ clusters) and is denoted as $R(t) \equiv \{c_1, c_2, \cdots, c_k\}$. Further, $k$ is determined as the value satisfying the inequalities $N_{k-1}(t) < \lfloor gN \rfloor \le N_k(t)$ for a given model parameter $g\in (0,1]$. $N_k(t) \equiv \sum_{\ell=1}^k s_{\ell}(t)$, where $s_{\ell}(t)$ is the number of nodes in the cluster $c_{\ell}$. We note that the number of clusters in $R(t)$ varies with the time step $t$. Here the time step $t$ is defined as the number of edges added to the system per node. 
This model is called a restricted ER model, because when $g=1$, the model is reduced to percolation in the ordinary ER model. We remark that this restricted ER model is a slightly modified version of the original model~\cite{half} in which the number of nodes in the set $R$ is fixed as $\lfloor gN \rfloor$. Thus, some nodes in a cluster on the boundary between the two sets $R$ and $R^{(c)}$ belong to the set $R$, and the others in the same cluster belong to the set $R^{(c)}$. However, for the modified model, all the nodes in the cluster are counted as elements of the set $R$. This modification enables one to solve the phase transition for $t > t_c$ analytically without changing any critical properties.
   
This restricted ER model exhibits an HPT at a transition point $t_c$. The order parameter $m(t)$, that is, the fraction of nodes belonging to the giant cluster, increases rapidly from zero at $t_c^{-}$ to a finite value $m_0$ at $t_c$. The interval $\Delta t=t_c-t_c^- \sim o(N)/N$. Thus, in the thermodynamic limit, this interval reduces to zero, and the order parameter is regarded as jumping discontinuously at $t_c$. For $t> t_c$, $m(t)$ increases gradually following formula~(\ref{eq:hpt_order}). Moreover, the size distribution of finite clusters, $n_s(t)$, exhibits power-law decay at $t_c$ with the exponent $\tau(g)$ in the range $2 < \tau(g) \le 2.5$. Thus, the critical exponents of the HPT vary continuously depending on the control parameter $g$. Such critical behaviors of the HPT in the cluster merging process have been observed for the first time.   
 
\section{Experimental results} 
 
 The EP models look too artificial, and one may wonder if the patterns produced by those models are physically relevant and can be observed experimentally in real-world systems. Along these lines, we introduce recent experimental results~\cite{enclave_nphy,enclave_prl}. The experiment was performed in a cytoskeletal system composed of actin filaments, fascin cross-links, and myosin motors. Actin filaments (bonds) and cross-links (sites) compose networks, and molecular motors exert localized stresses inside polymer networks to contract the crosslinked actin polymer network. As a result, small holes inside a large cluster are collapsed, and the large cluster becomes compact. Because of this compactness, the cluster size distribution exhibits power-law decay, but the exponent $\tau$ becomes less than two. Then, as we have already mentioned in Section 4, when the exponent $\tau$ is less than two, the percolation transition becomes discontinuous. The fact that the cluster becomes compact implies that the fractal dimension becomes two. Because the enclaving dynamic occurs suddenly, the order parameter jumps in a macroscopic scale within a short time interval.  
   
\section{Conclusion}

Since the paper on the EP model was published in 2009, a huge number of papers regarding this subject have been rapidly published. Thus, it is almost impossible to trace them all comprehensively. Here we have reviewed papers based on our publications ranging from the EP model to hybrid percolation models, and this review was written from our viewpoint. The subject of EP in which the exponent of the order parameter $\beta$ is extremely small or zero is still interesting, and many fundamental problems are not understood yet. More detailed reviews and open challenges in percolation can be found in Refs.~\cite{souza_nphy,epj}.

\ack
This work was supported by the National Research Foundation of Korea by grant no. NRF-2014R1A3A2069005. 
BK thanks the organization committee of the STATPHYS 26 conference for their invitation. This article is a transcription of an invited talk. He also thanks H.J. Herrmann, S. Hwang, J. Kert\'esz, D. Kim, J. Kim, J.S. Lee, M.G. Mazza, J. Nagler, and J. Park for their earlier collaborations on a subject related to this paper.

\end{document}